\documentstyle[12pt,axodraw]{article}
\begin{document}
\baselineskip=6.0mm
\begin{titlepage}
%%%%%%%%%%%%%%%%%%%%%%%%%%%%%%%%%%%%%%%%%%%%%%%%%
\begin{flushright}
  KOBE-TH-98-07\\
\end{flushright}
%%%%%%%%%%%%%%%%%%%%%%%%%%%%%%%%%%%%%%%%%%%%%%%%%%%
\vspace{1.0cm}
\centerline{{\large{\bf Grand Unified Theories}}}
\centerline{{\large{\bf and}}}
\centerline{{\large{\bf Lepton-Flavour Violation}}\footnote[1]{Talk given at 
the XVI Autumn School and Workshop on Fermion Masses, Mixing and CP 
Violation,
Institute Superior Tecnico, Lisboa, Portugal, 6-15 October 1997}}
\par
\par
\par\bigskip
\par\bigskip
\par\bigskip
\par\bigskip
\par\bigskip
\renewcommand{\thefootnote}{\fnsymbol{footnote}}
\centerline{{\bf C.S. Lim}\footnote[2]{e-mail:lim@oct.phys.kobe-u.ac.jp}}
\par
\par\bigskip
\par\bigskip
\centerline{Department of Physics, Kobe University, Nada, Kobe 657,
Japan}
\par
\par\bigskip
\par\bigskip
\par\bigskip
\par\bigskip
\par\bigskip
%\centerline{\today }\par
\par\bigskip
\par\bigskip
\par\bigskip
\par\bigskip
\centerline{{\bf Abstract}}\par

Lepton-flavour violating processes, such as $\mu \rightarrow e\gamma$,
 are studied in ordinary (non-SUSY) SU(5) and SUSY SU(5) grand unified
 theories. First given are some introductory argument on the mechanism
 of U.V. divergence cancellation in flavour changing neutral current
 processes and on the decoupling of particles with GUT scale masses
 . We next see that such general argument is confirmed by an explicit
 calculation of the amplitude of $\mu \rightarrow e\gamma$ in ordinary
 SU(5), which shows that logarithmic divergence really cancels among
 diagrams and remaining finite part are suppressed by at least
 $1/M_{GUT}^2$. In SUSY SU(5), flavour changing slepton mass-squared
 term get a logarithmic correction, as recently claimed. However, when
 the effect of flavour changing wave function renormalization is also
 taken into account such logarithmic correction turns out to
 disappear, provided a condition is met among SUSY breaking soft
 masses. In SUGRA-inspired SUSY GUT, such condition is not
 satisfied. But the remaining logarithmic effect is argued not to be
 taken as a prediction of the theory associated with an irrelevant
 operator. We find that the log-correction should not exist in models
 with gauge mediated SUSY breaking.

\par\bigskip
\par\bigskip
\par\bigskip
\par\bigskip
\par\bigskip
\par\bigskip
\par\bigskip

%\noindent September 1995
\end{titlepage}
\newpage

This paper is based upon the talk given at the XVI Autumn School and
Workshop on Fermion Masses, Mixing and CP Violation, Institute
Superior Tecnico, Lisboa, Portugal, 6-15 October 1997.  The former
part is aimed to give some introductory argument on the basic
properties of flavour changing neutral current processes, especially
on the mechanism of U.V. divergence cancellation and the decoupling of
heavy particles with GUT or SUSY breaking mass scales.  The latter
part is devoted to the lepton flavour violating processes in ordinary
(non-SUSY) SU(5) and SUSY SU(5), which is extensively based on the
recent original work with B. Taga \cite{Lim}.

\vspace{0.5 cm}
\leftline{\bf 1. Introduction}
\vspace{0.2 cm}

\vspace{0.3 cm}
\leftline{1.1 Flavour  violation and flavour changing neutral currents}
\vspace{0.2 cm}

So far we do not have any definitive argument which reveals the origin
of flavour. We do not know how quarks of the same charge have
different masses and how the mixing among these different flavours
stems.

Provided all quark masses are degenerate, the system of quarks has a
flavour symmetry, i.e. global ``horizontal symmetry" U(3) among three
generations.  The global symmetry leads to conserved charges, $N_d,
N_s,$ etc., `down number', strangeness, etc..  Hence flavour changing
neutral current (FCNC) processes, such as $s \rightarrow d\gamma$, are
strictly forbidden. For non-degenerate masses, the flavour symmetry
U(3) is broken, but there still remain U(1)$^3$ as the sub-symmetry,
which again forbids the FCNC processes.  The sub-symmetry is
eventually broken by flavour mixing described by the CKM mixing
matrix.  The violation of flavour symmetry (``flavour violation" for
short) now leads to FCNC.

FCNC processes have long served as good thinking ground and good
experimental probe of ``new physics" in each stage of the development.
They are caused in general by quantum effects of all possible
intermediate states, therefore being suitable places to search for
new unknown particles, predicted by new physics. They are also closely 
related with CP violation .
Typical examples to show these facts are the introduction of charm
quark by GIM \cite{Glashow},  the  following  prediction of the mass
by Gaillard and Lee by use of the rate of $K^0 \leftrightarrow
\overline{K^0}$ \cite{Gaillard}, and the Kobayashi-Maskawa
model \cite{Kobayashi} in order to explain CP violation in FCNC processes of 
neutral kaon system. Let us note that in the Kobayashi-Maskawa model any CP 
violating observable
should be handled by $det [M_u^{\dag}M_u, M_d^{\dag}M_d] = (m_u^2-m_c^2)
(m_c^2-m_t^2)(m_t^2-m_u^2)(m_d^2-m_s^2)(m_s^2-m_b^2)(m_b^2-m_d^2)\times J$, 
with $J$ being Jarlskog's parameter, which clearly suggests the
necessity of flavour violation or FCNC processes due to the violation
of U(3) by mass differences and flavour mixing.

Nowadays rare FCNC processes are still playing important roles in
searching for physics beyond the standard model, whose typical
examples are grand unified theories (GUT) and supersymmetric (SUSY)
theories. Flavour violation in the lepton sector, ``lepton flavour
violation (LFV)", is of special interest to us.  FCNC processes due to
LFV are strictly forbidden in the Standard Model, even though the
masses of charged leptons are all different. This is because neutrinos
are massless in the model; because of the absence of right-handed
neutrinos and lepton number conservation, neither Dirac-type nor
Majorana-type mass is possible.  For degenerate neutrino masses, in
the sense of $m_{\nu} = 0$, flavour mixing has no physical meaning, as
we can always perform a unitary transformation to move to the basis of
neutrino states where there is no flavour mixing.  Thus LFV is a very
clean signal of new physics, if it ever exists.  The typical examples
of LFV process are $\mu \rightarrow e\gamma$ and $\mu \rightarrow
ee\overline{e}$, where both of electron number and muon number, $N_e$
and $N_{\mu}$, change by one unit.  Our main interest here is $\mu
\rightarrow e\gamma$, whose experimental bound is $Br(\mu \rightarrow
e\gamma) < 4.9\times 10^{-11}$.

The standard model, by construction, has a nature that FCNC processes
are forbidden at the tree level, thanks to the idea of GIM.  In that
sense, we may say that FCNC is naturally suppressed in the standard
model. Let us note that when the differences of Yukawa couplings
vanish, the global flavour symmetry U(3) is enhanced in the theory ,
and therefore small FCNC is naturally preserved, according to the
argument of 't Hooft \cite{'tHooft}.  At the quantum level FCNC
processes are induced due to the flavour mixing in charged current
interaction handled by the CKM matrix.

\vspace{0.3 cm}
\leftline{1.2 Ultraviolet divergences and FCNC processes}
\vspace{0.2 cm}

As was discussed above, FCNC does not exist at the tree (classical) level. 
Generally speaking, if some observables are forbidden at the classical level 
(by some symmetery reason), the quantum corrections to them will be 
described in the form of effective lagrangian $L_{eff}$ with  higher 
dimensional ($d > 4$) gauge invariant operators, ``irrelevant operators", 
including Higgs field, in
general. (In gauge theories any  quantum  correction can be described by a 
gauge invariant operator when Higgs field is included, inspite of the 
spontaneous gauge symmetry
breaking.) The typical examples, besides FCNC, are the oblique corrections 
$S,T,U$ ($\epsilon_1$, $\epsilon_2$, $\epsilon_3$) \cite{Peskin}. For 
instance, the irrelevant operator, responsible for  the $T$-parameter, is
\begin{equation}
L_{eff}
=  C_{T} [(H^{\dag}D_{\mu}H)(H^{\dag}D^{\mu}H) - 
\frac{1}{3}(H^{\dag}D_{\mu}
D^{\mu}H)(H^{\dag}H)],
\end{equation}
where H denotes the Higgs doublet and $D_{\mu}$ stands for a covariant 
derivative.The coefficient function $C_{T}$ summalizes the quantum effects 
due to the intermediate states, and is related to the $T$-parameter as $T = 
- (\sin^{2}\theta_{W}M_W^2/ 3\pi\alpha)C_{T} $. The coefficient $C_{T}$ has 
a mass dimension $d = -2$, and
should be finite, as there is no counterterm in the original lagrangian to 
cancel the divergence even if
it appears while the theory is renormalizable.
In the same way as the case of $T$, the ultraviolet (U.V.) divergences in 
the quantum corrections to FCNC processes should  automatically disappear.

The above statement is rather trivial. But since the issue plays a central 
role in the forthcoming discussions, let us study the mechanism of the
divergence cancellation more carefully. Let us focus on the FCNC of quark
sector in the standard model. Diagrams can be devided into two
types depending on whether FCNC stems from ``soft" or ``hard" flavour 
violation.
In the (1-loop) diagrams with $W^{\pm}$ exchange, the flavour symmetry is
softly broken. This is because the gauge coupling is independent of flavour, 
and the only source to break the flavour symmetry is quark mass(-squared) in 
the quark propagator. The breaking is soft, in the sense that it is due to a 
quantity  with positive  mass dimension (quark mass(-squared)), and can be 
neglected
in higher energies (U.V. region). Hence this type of diagrams does not have 
any U.V. divergence. Or, one can directly confirm this by noting the 
unitarity
of CKM matrix, e.g.,
\begin{equation}
\sum_{i=u,c,t} V^{\dag}_{bi}V_{is} \ ln\Lambda = (V^{\dag}V)_{bs} \ 
ln\Lambda = 0 .
\end{equation}
 In the language of the gauge invariant operator, the quantum effects should 
be summalized by an  irrelevant operator ($d = 6$) with two Higgs doublets 
(to provide quark mass-squared after spontaneous symmetry breaking), whose 
coefficient should be finite. On the other hand in the diagrams where an 
unphysical scalar (``would be Nambu-Goldstone boson"), $\phi^{\pm}$, is 
exchanged, the flavour symmetry is hardly broken by the
dimensionless non-degenerate Yukawa-couplings of $\phi^{\pm}$ with quarks. 
Each of this type of diagrams, therefore, may give a contribution with U.V. 
divergence to a marginal ($d = 4$) operator. More precisely, both of flavour 
changing self-energy diagram (e.g. 
$\overline{s}i\partial_{\mu}\gamma^{\mu}b$) and flavour changing
vertex diagram (e.g. $\overline{s}\gamma^{\mu}b A_{\mu}$, with $A_{\mu}$ 
denoting photon field) have U.V. divergences. What we find \cite{Inami}, 
however, is that there is an exact cancellation  of
U.V. divergence between  the diagram where a photon is attached to the 
external legs of the  flavour changing self-energy diagram,  ``external leg 
correction" diagram shown in Fig.1 (a), (b),  and the proper diagram for 
flavour changing
vertex shown in Fig.1 (c), (d). Thus the final result is finite, in spite of 
the presence of the hard flavour violation.

%%%%%%%%%%%%%%%%%%%%%%%%%%%%%%%%%%%%%%%%%%%%%%%%%%%%%%%%%%%%%%%%%%%%%%%%%%%%%

\begin{center} 
\begin{picture}(350,60)(0,0)
\Line(5,40)(80,40)
\Line(95,40)(170,40)
\Line(185,40)(260,40)
\Line(275,60)(350,60)
%\Vertex(25,245){2.5}
\DashCArc(35,40)(15,0,180){2.5}
\DashCArc(140,40)(15,0,180){2.5}
\DashCArc(223,40)(20,0,180){2.5}
\DashCArc(313,60)(20,180,360){2.5}
\Photon(65,40)(65,10){3}{4}
\Photon(110,40)(110,10){3}{4}
\Photon(223,40)(223,10){3}{4}
\Photon(313,40)(313,10){3}{4}
\Text(10,45)[]{${b}$}
\Text(58,45)[]{${s}$}
\Text(73,45)[]{${s}$}
\Text(103,45)[]{${b}$}
\Text(118,45)[]{${b}$}
\Text(165,45)[]{${s}$}
\Text(195,45)[]{${b}$}
\Text(250,45)[]{${s}$}
\Text(285,55)[]{${b}$}
\Text(340,55)[]{${s}$}
\Text(35,48)[]{\small ${\phi^+}$}
\Text(140,48)[]{\small ${\phi^+}$}
\Text(223,53)[]{\small ${\phi^+}$}
\Text(313,48)[]{\small ${\phi^+}$}
\Text(55,25)[]{${\gamma}$}
\Text(120,25)[]{${\gamma}$}
\Text(230,25)[]{${\gamma}$}
\Text(320,25)[]{${\gamma}$}
\Text(42,2)[]{\sl (a)}
\Text(132,2)[]{\sl (b)}
\Text(222,2)[]{\sl (c)}
\Text(312,2)[]{\sl (d)}
%%%
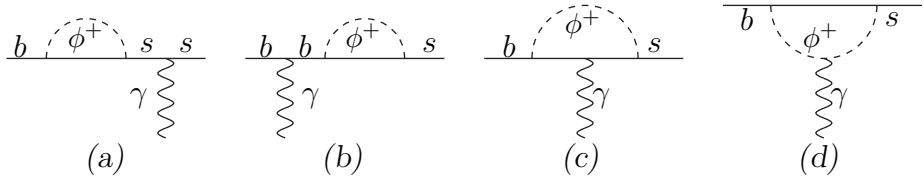
\end{picture} \\{{\sl Fig.1 }{\rm : The diagrams with logarithmic divergences contributing to the amplitude of ${d \to s \gamma}$ transition in the standard model. (a), (b) are called external leg corrections and (c), (d) are proper diagrams for flavour changing vertex. ${\phi^\dagger}$ denotes a would-be N.-G. boson. }} 
\end{center}
%%%%%%%%%%%%%%%%%%%%%%%%%%%%%%%%%%%%%%%%%%%%%%%%%%%%%%%%%%%%%%%%%%%%%%%%%%%%%

One may feel uncomfortable with the diagrams where a photon is
attached to the external legs of the flavour changing self-energy
diagram, since these diagrams are 1-particle reducible. If we wish we
may add the contribution of counterterm for flavour changing
self-energy, so that the external leg correction disappears. The net
effect of flavour changing counterterm contributions, however, turns
out to vanish exactly \cite{Botella}.  Let $\psi_0 $ and $\psi $ be
the column vectors of bare and renormalized down type quarks,
respectively: $\psi_0^t = (d_0, s_0, ...) $ and $\psi^t = (d, s, ...) 
$. They are related not only by rescaling but also by unitary
transformation among generations, which is necessary to get flavour
changing couterterms; $\psi_0 = (Z_L \frac{1 - \gamma_5}{2} + Z_R
\frac{1 + \gamma_5}{2})\psi $ where renormalization constants $Z_L $
and $Z_ R $ for each chirality are $ 3 \times 3 $ matrices.  The
obtained counterterms are written as
\begin{equation}
L_c = \overline{\psi}i\partial_{\mu}\gamma^{\mu}(A\frac{1 - 
\gamma_5}{2}+B\frac{1 + \gamma_5}{2})\psi
- \frac{e}{3}\overline{\psi}\gamma_{\mu}(A\frac{1 - \gamma_5}{2}+B\frac{1 + 
\gamma_5}{2})\psi A^{\mu},
\end{equation}
where $A = Z_L^{\dag}Z_L - I, B = Z_R^{\dag}Z_R - I$. Let us note that the 
wave function
renormalization for the photon field $A^{\mu}$, $Z_3$, does not contribute 
to flavour changing counterterm (at the 1-loop level),
and the counterterms for the quark self-energy and photon vertex are not 
independent (Ward identity, $Z_1 = Z_2$). This Ward identity is responsible 
for the vanishing net contribution of
the counterterms, irrespectively of the choice of $A$ and $B$ 
\cite{Botella}.
In other words, we may say that in terms of the renormalized field $\psi$, 
defined
such that their kinetic terms are flavour diagonal, the flavour changing 
vertex correction should be
finite. If the flavour changing self-energy counterterm is chosen so that it 
cancels the
corresponding 1-loop contribution, the external leg correction just 
vanishes. In terms of such defined renormalized field $\psi$, the sum of the 
1-loop proper diagram  for vertex correction and the contribution of the 
counterterm for flavour changing vertex should be finite, as
the sum of all 1-loop diagrams and counterterm contributions should be 
finite.

We may summalize this situation relying on operator language as follows.
Because of the $U(1)_{em}$ invariance, after the quantum corrections, the 
relevant
marginal operator can be written in terms of the bare fields $\psi_{0}$ and 
 a covariant derivative $D_{\mu}$ as,
\begin{equation}
\overline{\psi_0}iD_{\mu}\gamma^{\mu} H \psi_0,
\end{equation}
where the $ 3 \times 3 $ hermitina matrix $ H $ is generally allowed to have 
flavour changing off-diagonal elements.
Hence, even if the original lagrangian is assumed to be flavour-diagonal, 
after the
quantum correction the effective lagrangian is no longer flavour diagonal.
However, we may move to the basis of the renormalized fields $\psi$ so that 
their kinetic term becomes flavour diagonal, $U^{\dag}HU = H_{diagonal}$. 
This unitary
transformation, at the same time, diagonalizes the photon vertex coming from 
 $D_{\mu}$. We may further perform the rescaling of the fields
so that the kinetic term is proportional to $I$, a unit matrix. Then another 
unitary transformation becomes possible to make the mass term of the quarks 
flavour-diagonal, while keeping the form of the kinetic term.
Thus again FCNC disappers from the whole relevant or marginal operator. 
There remain flavour changing irrelevant
operators, but each of them is accompanied by finite coefficient. In fact, 
 this freedom of unitary transformation and rescaling makes it possible to 
assume
always that the original lagrangian to start with is flavour diagonal.
The lesson here is the importance of redifinition or wave function 
renormalization of matter fields caused by the
flavour changing self energies.

\vspace{0.3 cm}
\leftline{1.3 Decoupling}
\vspace{0.2 cm}

Another issue which also will play a central role in the later discussion is 
the concept of
decoupling. Theories beyond the standard model, ``new physics",  generally 
predict the presence of unkown heavy
particles. Then an important question is whether their presence can be 
searched for in low
energies, $E \ll M$,  with $M$ being a generic heavy particle mass. 
Concerning this, there is a
well-known decoupling theorem by Appelquist and Carazzone \cite{Appelquist}. 
The theorem is
valid in vector-like (non-chiral) gauge theories without spontaneous gauge 
symmetry breaking,
whose typical examples are QED and QCD. The theorem says that the effects of 
a heavy particle
with mass $M$ in low energy processes are suppressed by the inverse power of 
$M$. More precisely,
the quantum effects of the heavy particle may affect the coefficients of 
relevant or marginal
operators, having mass dimension $d \leq 4$, behaving as $M^2$ or $ln M$. 
But they just affect
the relation between the bare and renormalized quantities, i.e. 
renormalization. On the other hand,
quantum effects to irrelevant operators are genuine predictions of the 
theory. However, the
coefficient functions, having $d < 0$, are suprresed by the power of $1/M$. 
Fortunately, in chiral theories with spontaneous gauge symmetry breaking, 
whose typical example is
the standard model, there are a few observables known where the 
contributions of heavy particles are not suppressed:
non-decoupling. Good example will be the t quark contributions bahaving as 
$m_t^2$ to the $\rho$-parameter \cite{Veltman} or to the FCNC processes 
\cite{Inami}  in the standard model.

Whether heavy particles decouple from low-energy processes or not depends on 
the nature of
new physics, more specifically on the origin of the large mass scale $M$.

(i) The first to be considered is the case where $M$ is provided by a new 
large ($\gg M_W$)
 $SU(3) \times SU(2) \times U(1)$ singlet mass scale $M_s$. For instance, in 
SUSY theories $M_s$ should be
the SUSY breaking scale $M_{SUSY}$, and in GUT theories $M_s$ shoukd be
taken to be $M_{GUT}$. In this case the contributions of heavy particles, 
such as
super-partners or particles with GUT scale masses, are decoupled. The 
quantum effects of heavy particles are summalized
by $L_{eff} = \sum_{} C_{i}O_{i}$, where $O_{i}$ are gauge invariant 
irrelevant operators with dimension
$d_i > 4$. The contribution of a heavy particle to the gauge invariant 
coefficient $C_i$
should behave as $C_i \propto 1/M_{s}^{d_i-4}$, leading to the decoupling. 
One example is the contribution of super-partners of light quarks, the 
doublet of first generation
 $(\tilde{u},\tilde{d})^t$, to the $\rho$-parameter. Writing the squark 
masses as $m_{\tilde{u},\tilde{d}}^2 = m_{u,d}^2 + M_{SUSY}^2$, we find
for $m_{u,d}^2 \ll M_{SUSY}^2$ (neglecting the left-right squak mixings for 
brevity) \cite{Lim2},
\begin{equation}
\Delta\rho \simeq \frac{g^2}{(4\pi)^2}\frac{1}{4}\frac{(m_u^2 - 
m_d^2)^2}{M_W^{2}M_{SUSY}^2}.
\end{equation}
This result clearly shows the suppression by $1/M_{SUSY}^2$.

(ii) The second is the case where $M$ is provided by a large coupling with 
Higgs scalars,
through spontaneous symmetry breaking. In this case there is no new mass 
scale, i.e. $M_s \sim M_W$, and
non-decoupling phenomena are expected. One example is found in the 
contribution of t quark to a FCNC process, $B_{d} - \overline{B_{d}}$
mixing. The relevant effective lagrangian is given as
\begin{equation}
L_{eff} = \frac{G_F}{\sqrt{2}}\frac{\alpha}{4\pi 
sin^2\theta_{W}}(V_{bt}^{\dag}V_{td})^2 E(x)
 (\overline{b}_L \gamma_{\mu} d_L)^2  .
\end{equation}
where $x = m_t^2/ M_W^2$. For large $m_t$, $E(x)$ behaves as $(-1/4)x = 
(-1/4) (m_t^2/ M_W^2)$,
thus leading to a non-decoupling effect of t quark \cite{Inami}.
The factor $m_t^2/ M_W^2$ may be understood to come from a factor 
$f_t^4/m_t^2
\sim g^{4}m_t^{2}/M_{W}^4$, with $f_t$ being the top Yukawa coupling 
 appearing in the box diagram with $\phi^{\pm}$ exchange.

\vspace{0.3 cm}
\leftline{1.4 Lepton flavour violation and grand unified theories}
\vspace{0.2 cm}

In the standard model lepton flavour violation (LFV) is strictly
forbidden, as we have already seen. The situation is quite different when 
the standard model is embedded in
GUT-type theories. The essential difference stems from the fact that in GUT 
there exist interactions which connect leptons with quarks through the
exchange of leptoquark particles, and that flavour symmetry is broken
in the quark sector. Then the difference between unitary transformations in 
charged lepton and quark sectors will result in LFV processes, such as
$\mu \rightarrow e\gamma $.
We, however, expect that the rate of such LFV processes are extraordinarily 
small, being suppressed by a factor $(\Delta m_{q}^2/ M_{GUT}^2)^2 $, where 
$\Delta m_{q}^2$  denotes mass-squared differnces of quarks.
This suppression is suggested by the argument for the decoupling of GUT 
particles. Such suppression by
$1/ M_{GUT}^4$ is similar to what happens in the proton decay.

\vspace{0.3 cm}
\leftline{1.5 Lepton flavour violation in supersymmetric grand unified 
theories}\vspace{0.2 cm}

Next we will focus on LFV in supersymmetric grand unified theories (SUSY 
GUT),  which is  of great current interest.
What is new in SUSY theories concerning LFV ? There seems to be two new 
sources of LFV even in ordinary
SUSY models without grand unification, which will be discussed below. \\
(a) ``Sizable" (larger than we usually expect) rates of LFV become possible, 
once R-invariance  is relaxed. Unless we impose R-invariance, superpotential 
$W$ may have new
source of flavour violation,
\begin{equation}
W = c_1 \ \overline{u}\overline{d}\overline{s} + c_2 \ e u \overline{d} + 
c_3 \  \mu u \overline{d} + .... \ \ ,
\end{equation}
in addition to ordinary R-invariant couplings of Higgs scalar. The term with 
coefficient $c_1$ violates baryon number and strangeness, $|\Delta B| = 
|\Delta s| = 1$. The terms with coefficients $c_2$ and $c_3$ violate lepton 
number and electron or muon number,  $|\Delta L| = 1$, $|\Delta N_e|$ or 
$|\Delta N_{\mu}| = 1$. Thus the
combined effect of these couplings yields $\mu \rightarrow e\gamma$
via $\tilde{u}$-exchange, for instance.   \\
(b) Another possible source of LFV is soft SUSY breaking terms. For 
instance,
gauge invariant LFV  slepton mass term,
\begin{equation}
M_{SUSY}^2 \ \tilde{e}_R^{\ast} \tilde{\mu}_R ,
\end{equation}
leads to $\Delta N_e = -\Delta N_{\mu} = 1$ processes, including $\mu 
\rightarrow e\gamma$.
 However, in supergravity (SUGRA) inspired MSSM R-invariance is assumed and 
the soft breaking terms themselves induced by SUGRA interactions
are assumed to be universal and not to break flavour symmetry at least at 
the tree level. Thus these new features of SUSY theories seem not to play a 
crucial role in LFV processes.

In this context, a very interesting claim has been made by Barbieri - Hall 
and collaborators that in SUSY GUT models ``sizable" rates of $\mu 
\rightarrow e\gamma$ are
expected, contrary to ordinary expectation \cite{Barbieri}. Such claim has 
been followed by many works calculating the rate in a few GUT theories
\cite{Muegamma}. The crucial observation there is that even in R-conserving 
SUGRA-inspired SUSY GUT models, with flavour symmetric universal SUSY 
breaking
slepton masses, the large flavour violation due to top quark Yukawa coupling 
combined  with GUT interaction give sizable non-universal or LFV 
renormalization group effects on the SUSY breaking slepton masses. They 
cause the discrepancy between mass
matrices of charged leptons and their superpartners. Thus super-GIM 
mechanism
\cite{Gatto} is no longer valid and
the resultant term in Eq.(8) and photino-exchange, for instance, leads to 
$\mu \rightarrow
e\gamma$ (one of the possible daigrams is shown in Fig.2), at the rates 
which is not  so far from the present experimental upper bound .

%%%%%%%%%%%%%%%%%%%%%%%%%%%%%%%%%%%%%%%%%%%%%%%%%%%%%%%%%%%%%%%%%%%%%%%%%%%%%
\begin{center} 
\begin{picture}(100,50)(0,0)
%%%
\Line(30,25)(5,25) \Line(70,25)(95,25) \CArc(50,25)(20,0,180)
\CArc(50,25)(3,0,360)
%\Vertex(45,50){2.5}
\Line(48,27)(52,23)
\Line(48,23)(52,27)
\Photon(60,25)(78,1){3}{4}
\DashLine(30,25)(47,25){2.5}
\DashLine(53,25)(70,25){2.5}
\Text(15,30)[]{${\mu}$}
\Text(40,30)[]{${\tilde{\mu}}$}
\Text(60,30)[]{${\tilde{e}}$}
\Text(85,30)[]{${e}$}
\Text(50,38)[]{${\tilde{\gamma}}$}
\Text(83,1)[]{${\gamma}$}
%%%
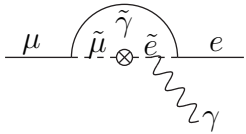
\end{picture} \\{{\sl Fig.2 }{\rm : A 1-loop diagram for ${\mu \to e \gamma}$ decay, caused by the lepton flavour changing mass-squared term denoted by a blob. }}
\end{center} 
%%%%%%%%%%%%%%%%%%%%%%%%%%%%%%%%%%%%%%%%%%%%%%%%%%%%%%%%%%%%%%%%%%%%%%%%%%%%%

While their result \cite{Barbieri}is quite impressive, it seems to be
different from what we expect according to our argument given
above. Namely, \\ (i) The reason they got sizable rates is the
presence of aforementioned LFV slepton masses due to renormalization,
which are proportional to $ln \frac{\Lambda}{M_{GUT}}$, instead of
$\frac{m_{\mu}^{2}}{M_{GUT}^2}$. The difference is outrageous ! The
U.V. cutoff $\Lambda$ was taken to be the Planck scale $\Lambda =
M_{pl}$.  This, however, in turn mean that logarithmic divergence
remains in the quantum effects on LFV slepton masses in SUSY GUT, and
seems to contradict with what we have seen above in the standard model
and what we expect in non-SUSY SU(5) discussed below . \\ (ii) Next,
the $ln \frac{\Lambda}{M_{GUT}}$ contribution clearly shows that
GUT particles do not decouple from the low energy process, in contradiciton 
with our analysis based on general argument that GUT particle contributions 
should decouple.

The purpose of our work \cite{Lim} is to clarify whether the interesting 
features claimed in Ref.\cite{Barbieri}
 are natural consequences of SUSY GUT theories. We will compare LFV in 
ordinary non-SUSY
SU(5) GUT and in SUSY SU(5) GUT, in order to see how SUSY can be essential 
in getting the sizable effects. A special attention will be paid on the
effect of lepton-flavour changing wave function renormalization, which 
played
a central role in the cancellation of U.V. divergences in the standard 
model.

\vspace{0.5 cm}
\leftline{\bf 2. Lepton flavour violation in ordinary non-SUSY SU(5) GUT}
\vspace{0.2 cm}

The content of this section is based on the master thesis of B.Taga 
\cite{Taga}.  As we have discussed in the introduction, even if neutrinos 
are massless
GUT interactins connecting leptons with quarks make LFV possible, though the 
rates are expected to be quite strongly suppressed by $1/M_{GUT}^4$.
The purpose here is to confirm this expectation by actual calculations.
As LFV becomes possible due to GUT interactions, we may focus only on the
diagrams with exchanges of heavy particles with GUT mass scale $M_{GUT}$. 
More precisely
the exchanges of $Y_{\mu}$ (lepto-quark gauge boson which connects charged 
leptons and up-type quarks), $h^{\prime}$ (color triplet Higgs ) and 
$\Sigma_{Y}^{\prime}$ (Nambu-Goldstone mode for $Y_{\mu}$). (The prime of 
$h^{\prime}$ and $\Sigma_{Y}^{\prime}$ denotes a small admixture of  color 
triplet components of 5-plet and adjoint Higgs fields) . One remark here is 
that another lepto-quark gauge boson $X_{\mu}$, which connects charged 
leptons with down-type quarks, does not
contributes to the LFV. This is simply because in the GUT scheme the mass
matrices of charged lepton and of down-type quarks are the same (at tree 
level).We find that out of 14 Feynman diagrams, only $h^{\prime}$-exchange 
diagrams give numerically important contributions. Let $L_{eff}$ be an 
effective lagrangian, which is responsible for the $\mu \rightarrow e\gamma$ 
decay,
\begin{equation}
L_{eff} = c_{LFV}\times \overline{e}\sigma_{\mu\nu}(m_{\mu}\frac{1 - 
\gamma_5}{2} +
m_{e}\frac{1 + \gamma_5}{2}) \mu \times F^{\mu\nu},
\end{equation}
where $ F^{\mu\nu}$ is the photon field strength. The contribution of each 
type of diagrams to the coefficient function $c_{LFV}$  is given as
\begin{equation}
c_{LFV}(h^{\prime}-\mbox{exchange}) = - 
\frac{\sqrt{6}}{8}\frac{g^3}{(4\pi)^2}
\frac{1}{M_h^2}(V_{KM}^{\dag})_{\mu 
j}(V_{KM})_{je}\frac{m_{uj}^2}{M_W^2}(2ln\frac{m_{uj}^2}{M_h^2} + 
\frac{15}{4}),
\end{equation}
\begin{equation}
c_{LFV}(Y_{\mu}, \Sigma_{Y}^{\prime}-\mbox{exchange}) = 
O(g^2\frac{m_{uj}^2}{M_{Y}^4})(V_{KM}^{\dag})_{\mu j}(V_{KM})_{je},
\end{equation}
where $M_h$ denotes the mass of $h^{\prime}$ and the Kobayashi-Maskawa 
matrix $V_{KM}$ handles the LFV as well. The contributions of $Y_{\mu}$ and 
$\Sigma_{Y}^{\prime}$-exchange diagrams are indistinguishable, as the gauge 
invariance is
guaranteed when they are  summed up. Thus only the 
$c_{LFV}(h^{\prime}$-exchange) is important, though the resultant branching 
ratio is almost nothing being suppressed by $1/M_{h}^4 \sim 1/M_{GUT}^4$. 
The difference of the suppression factors in $c_{LFV}(h^{\prime}$-exchange) 
and $c_{LFV}(Y_{\mu}, \Sigma_{Y}^{\prime}$-exchange) is understood as 
follows. In QED, $\mu \rightarrow e\gamma$
is described by $d = 5$ operator $\overline{e}\sigma_{\mu\nu} \mu \times 
F^{\mu\nu}$. In the standard model or GUT this operator cannot be gauge 
invariant
as $e_L$ and $\mu_R$ have different quantum numbers. Thus a Higgs (doublet 
or 5-plet)
should be included into the operator, which makes the dimension of the 
operator $d = 6$. This is why $c_{LFV}(h^{\prime}$-exchange) is suppressed 
by $1/M_{GUT}^2$. In the case of $c_{LFV}(Y_{\mu}, 
\Sigma_{Y}^{\prime}$-exchange) we
need 2 more Higgs fields, as lepton flavour is only softly broken by the
insertion of quark masses twice, thus making the operator $d = 8$. The 
coefficient is thus suppressed by $1/M_{GUT}^4$.

We have shown two things by explicit calculation, i.e., (i) the 
log-divergence cancells out when the sum of all $h^{\prime}$-exchange 
diagrams is taken, and
 (ii) the decoupling of GUT particles also holds; after the cancellation of 
the log-divergence and constant terms, the remaining amplitude is 
 suppressed at least by $1/M_{GUT}^2$. (Actually, the log-divergence 
cancellation has a relevance  only for the chirality preserving `charge 
radius' term, not for the `magnetic
moment' term responsible for  $\mu \rightarrow e\gamma$ at 1-loop level. Let 
us note, however, that what plays a central role in  $\mu \rightarrow 
e\gamma$ in SUSY SU(5) is chiralty preserving flavour changing right-handed 
slepton mass-squared term.)

\vspace{0.5 cm}
\leftline{\bf 3. Lepton flavour violation in supersymmetric SU(5) GUT}
\vspace{0.2 cm}

In Ref.\cite{Barbieri} it has been claimed that the rate of $\mu \rightarrow 
e\gamma$
can be `sizable' in SUSY GUT. We, however, have already seen in the previous 
section that such sizable effects
do not appear in ordinary non-SUSY SU(5) theory. Why $\mu \rightarrow 
e\gamma$ is
enhanced when the theory is made supersymmetric ? Conceptually, SUSY
itself is
independent of flavour symmetry.
More explicitly, we have the following questions raised concerning their 
results; \\
(i) Why does the logarithmic-divergence $ln \frac{\Lambda^2}{M_{GUT}^2} = ln 
\frac{M_{pl}^2}{M_{GUT}^2}$, implied by the claimed renormalization group 
effect, remain ? \\
(ii)Why is the suppression by a factor $(\frac{M_{W}^4}{M_{GUT}^4})$, i.e. 
the decoupling, absent ? \\
The Taga's thesis \cite{Taga} has shown that these are not the case in 
non-SUSY SU(5) GUT.

Though we know that in SUSY GUT the soft SUSY breaking masses can be
new source of flavour violation, we still have the following questions; \\ 
(iii) In Ref.\cite{Barbieri}, soft SUSY breaking masses have been assumed to 
be universal, being flavour independent at the tree level and cannot be a 
new source of LFV. Then why did such drastic change as 
$(\frac{M_{W}^4}{M_{GUT}^4})$ $\rightarrow ln \frac{M_{pl}^2}{M_{GUT}^2}$ 
become possible ? \\
(iv) SUSY breaking terms are quite soft, i.e. $M_{SUSY} \ll M_{GUT}, \ 
M_{pl}= \Lambda$. Then how can they drastically affect the U.V.-divergence ?

  Since LFV in SUSY GUT is so interesting and important issue, to settle the 
above questions seems to be quite meaningful. We therefore try to reanalyze 
the issue with the help of concrete computations. Our main interest is in 
the point whether or not the logarithmic
divergence naturally remains and whether the general argument for the 
decoupling of the effects due to  GUT interactions really breaks down.
The model to work with is SUSY SU(5) GUT with explicit soft SUSY breaking 
terms.
According to Ref.\cite{Barbieri}, we assume that bare SUSY breaking masses, 
the masses at the cutoff $\Lambda$, are flavour-independent.
In particular, for right-handed charged sleptons,  they are given as
\begin{equation}
m_{0l}^2 (|\tilde{e}_R|^2 + |\tilde{\mu}_R|^2 + ...),
\end{equation}
where $m_{0l}$ denotes a universal bare SUSY breaking mass. Namely SUSY 
breaking itself cannot be the source of LFV. In MSSM, therefore, such 
flavour-independent masses will enable us to diagonalize both lepton and 
slepton mass matrices
simultaneously (super GIM-mechanism \cite{Gatto}), and for massless 
neutrinos there will be
no LFV, just as  in the standard model. As was pointed out by Barbieri-Hall 
and  collaborators, the situation changes in SUSY SU(5),
because of the presence of GUT interactions. The GUT interaction which 
connects charged sleptons with stop $\tilde{t}$, accompanied by large top 
Yukawa coupling $f_t$, can be new source of LFV. When combined with SUSY 
breaking, such effect leads to    LFV SUSY breaking masses like $(f_t^2 ln 
\frac{\Lambda^2}{M_{GUT}^2} M_{SUSY}^2) \ \tilde{e}_R^{\ast}\tilde{\mu}_R$. 
Let us note that this operator is a relevant operator having
$ d=2 $, and therefore the appearance of the logarithmic divergence just 
signals that such operator was possible to exist
in the original lagrangian, though it was not included in our choice.   The 
induced LFV slepton mass-squared term  may make super-GIM mecahnism invalid 
and may lead to $\mu \rightarrow e\gamma$
through the ordinary MSSM interactions, for instance photino-exchange 
diagram, whose amplitude is suppressed
just by the power of $1/M_{SUSY}^2$, not by $1/M_{GUT}^4$ . Thus actually in 
this scenario the $\mu \rightarrow e\gamma$ process is induced by the 2-loop 
effect. Let us note that if SUSY was not
broken LFV will not get sizable effects, just as shown in Taga's thesis
\cite{Taga} in
non-SUSY SU(5) theory. The combined effects of GUT and SUSY breaking is 
crucial \cite{Barbieri}.

The most rigorous way to reach the rate of $\mu \rightarrow e\gamma$ is to 
calculate all possible 2-loop diagrams,
which we would like to avoid. Instead, we may take the
following approach to analyze the effect. First, we perform the 
path-integral
from the cutoff $\Lambda$ to some scale $\mu$ , which satisfies  $M_W \ll 
\mu
\ll M_{GUT}$ (``Wilsonian renormalization"), at the 1-loop level. We thus 
obtain SU(3)$\times$SU(2)$\times$U(1) invariant effective low-energy ( $E 
\leq \mu$ ) lagrangian $L_{eff}$ with respect to light ($\ll M_{GUT}$) 
particles, which should be
identified with MSSM. Since $ \mu
\ll M_{GUT}$, even if we let heavy GUT particles remain in $L_{eff}$, their 
effects in low energies are suppressed by
at least $\mu^2/M_{GUT}^2$ anyway,  and these particles can be safely 
neglected in  $L_{eff}$. Then, by using the induced LFV masses for charged 
sleptons
in $L_{eff}$ the rate of $\mu \rightarrow e\gamma$  can be calculated just 
as in MSSM.  Thus our focus is on the point whether the non-decoupling 
logarithmic
LFV quantum correction ever appears in $L_{eff}$.
The effective lagrangian can be decomposed into $L_{eff} = L_{rel}
+ L_{irrel}$, where $L_{rel}$ includes operators with $d \le 4$, while
$L_{irrel}$ denotes the set of irrelevant operators with $d > 4$. Since LFV 
stems only from GUT particle exchanges, only their contributions to the LFV 
parts of $L_{eff}$ are considered below. Some remarks are in order; \\
(a) Even if we get flavour changing slepton masses by the loop integral of 
the proper diagrams, it does not immediately lead to the presence of FCNC. 
As
we have discussed in the introduction, the effects of flavour changing 
wave-function renormalization (self-energy)  should also be taken into 
accounts. This point seems to have been missed in
the previous analysis \cite{Barbieri}, \cite{Muegamma}. \\
(b)  The GUT particle contributions to $L_{irrel}$ will be suppressed by the 
inverse powers of $M_{GUT}$. Thus only the contributions to $L_{rel}$ will 
be
considered below. \\
(c) Although they are ``soft", the SUSY breaking terms potentially affect 
operators with
$d \leq 3$ in $L_{rel}$, but only up to $O(M_{SUSY}^2)$. For instance, the 
$M_{SUSY}^2$ insertion to the 1-loop diagram for LFV mass operator 
$\tilde{e}^{\ast}\tilde{\mu}$ yields
log-divergence, which plays important role in the discussion of 
Ref.\cite{Barbieri}. One more insertion of $M_{SUSY}^2$ will make the 
diagram finite, being suppressed by the inverse powers of $M_{GUT}$ (If we 
regard $M_{SUSY}^2$ as due to
the VEV of some spurious superfield's F-component, the corresponding 
operator
actually may be regarded as a irrelevant one). \\
(d) To get the flavour changing slepton masses, not only $M_{SUSY}^2$
but also  flavour violation are necessary. Only hard flavour violation due 
to the Yukawa  couplings via the
exchange of colored Higgs superfield will be important. The soft flavour 
violation due to the  mass-squared differences of up-type quarks
contributes to the process only as a irrelevant operator, since $M_{SUSY}^2$ 
insertion together with the insertion of Higgs doublet twice to provide the 
up-quark masses makes the operator higher-dimensional. Thus the effects of 
the soft flavour violation will be suppressed by the inverse of 
$M_{GUT}^2$.

Hence our task is to calculate the quantum corrections due to the exchange 
of colored Higgs superfield to the slepton part of $L_{rel}$ ( The quantum 
corrections to charged leptons are not independent of the supersymmetric 
terms of the slepton part, both being diagonalized symultaneously. Thus we 
need not calculate them independently). The relevant part of the calculated 
$L_{eff}$ (in momentum space) takes the following form;
\begin{eqnarray}
L_{eff}
&=&  \tilde{l}_i^{\ast}(p) [ (\delta_{ij} + a H_{ij} ) p^2 - (\delta_{ij} + 
b H_{ij}) m_{0l}^2 ] \tilde{l}_j(p)  \nonumber \\
&-&e \tilde{l}_i^{\ast}(p) (p + p')_{\mu} (\delta_{ij} + c H_{ij} ) 
\tilde{l}_j(p') \cdot A^{\mu}(q),
\end{eqnarray}
where $\tilde{l}_i = \tilde{e}_R, \tilde{\mu}_R$ etc., and $q = p - p'$. 
$A^{\mu}$ denotes photon field. The $m_{0l}$ is the SUSY breaking mass for 
slepton, and $H_{ij} = f_{ti}f_{tj}^{\ast}$ with $ f_{t\mu} = 
g(\frac{m_t}{M_W})V_{ts}^{KM}$, for instance. Let us note that the large top 
Yukawa coupling gives contributions only to the
operators with respect to right-handed sleptons. Thus terms in the above 
equation are all chirality preserving, and the SUSY breaking mass-squared 
term with $m_{0l}^2$ should be treated on an equal footing with the self 
energy
term accompanied by $p^2$.
The coefficients $a,b$, and $c$ contain the results of 1-loop calculations. 
 The log-divergent parts of these coefficients, of our main interest,
are given as,
\begin{eqnarray}
a &=&  -3 \Delta  \nonumber \\
b &=& 3 \cdot \frac{m_{0u}^2+m_{0h}^2+m_{03}^2}{m_{0l}^2}\cdot \Delta
\nonumber \\
c &=& a,
\end{eqnarray}
where $ \Delta = i \int \frac{d^{4}k}{(2\pi)^4}\frac{1}{(k^2+M_{GUT}^2)^2}$ 
with a generic GUT mass scale $M_{GUT}$. $m_{0l}^2$, $m_{0u}^2$, $m_{0h}^2$ 
are SUSY breaking masses for right-handed charged slepton, right-handed 
up-type squarks, and 5-plet Higgs, respectively, and $m_{03}$ is SUSY 
breaking trilinear coupling of
$\tilde{l}\tilde{t}h$. (As far as the log-divergence is concerned, Wilsonian 
renormalization and full integration do not make any difference.) The third 
relation $c = a$ is the consequence of U(1)$_{em}$ symmetry, or Ward
identity, as discussed in the introduction.

%%%%%%%%%%%%%%%%%%%%%%%%%%%%%%%%%%%%%%%%%%%%%%%%%%%
\begin{center} 
\begin{picture}(300,70)(0,0)
%%%
\DashLine(5,35)(32,35){2.5}
\DashLine(38,35)(62,35){2.5}
\DashLine(68,35)(95,35){2.5}
\DashLine(115,35)(142,35){2.5}
\DashLine(148,35)(195,35){2.5}
\Text(215,35)[l]{${\ldots \ldots }$}
%\DashLine(215,35)(275,35){2.5}
\Text(105,35)[]{${+}$}
\Text(205,35)[]{${+}$}
\Text(285,35)[]{${=}$}
\Text(300,35)[]{\Large ${0}$}
%\Photon(280,25)(305,25){1}{5}
%\Photon(280,23)(305,23){1}{5}
\CArc(35,35)(3,0,360)
\Line(33,37)(37,33)
\Line(33,33)(37,37)
\CArc(65,35)(3,0,360)
\Line(63,37)(67,33)
\Line(63,33)(67,37)
\CArc(145,35)(3,0,360)
\Line(143,37)(147,33)
\Line(143,33)(147,37)
\Photon(80,35)(80,5){3}{4}
\Photon(170,35)(170,5){3}{4}
\Text(15,40)[]{${\tilde{\mu}}$}
\Text(50,40)[]{${\tilde{e}}$}
\Text(75,40)[]{${\tilde{e}}$}
\Text(85,40)[]{${\tilde{e}}$}
\Text(125,40)[]{${\tilde{\mu}}$}
\Text(160,40)[]{${\tilde{e}}$}
\Text(185,40)[]{${\tilde{e}}$}
\Text(35,25)[]{${a p^2}$}
\Text(65,25)[]{${m^2_{0L}}$}
\Text(145,25)[]{${b m^2_{0L}}$}
%\Text(245,42)[]{(?)}
\Text(90,20)[]{${\gamma}$}
\Text(180,20)[]{${\gamma}$}
\Text(50,2)[]{\sl (a)}
\Text(155,2)[]{\sl (b)}
%\Text(245,2)[]{\sl (c)}
%%%

\end{picture} \\{{\sl Fig.3 }{\rm : Diagrams contributing to the sub-process ${\tilde{\mu} \to e \tilde{\gamma}}$ of ${\mu \to e \gamma}$ at ${{\cal O} \left( M^2_{\rm SUSY} \right)}$. When $a = b$ holds the sum of these diagrams vanishes. }} 
\end{center} 
%%%%%%%%%%%%%%%%%%%%%%%%%%%%%%%%%%%%%%%%%%%%%%%%%%%%%%%%%%%%%%%%%%%%%%%%%%%

When a specific relation $a = b = c$, which is equivalent to
\begin{equation}
m_{0l}^2 + m_{0u}^2 + m_{0h}^2 + m_{03}^2 = 0,
\end{equation}
holds, the SUSY breaking term $(\delta_{ij} + b H_{ij}) m_{0l}^2 $
has the same structure as other terms with coefficients $a$ and $c$, and 
these terms (and the kinetic term
for charged leptons) can be diagonalized and rescaled simultaneously by a 
suitable wave-function
renormalization. Thus  all LFV effects go away in $L_{rel}$, and the 
log-divergence $\Delta$ does not remain. Let us note the photino
vertex does not have any LFV either, as both lepton and slepton mass 
matrices
are simultaneously diagonalized.  This disappearance of LFV can also be 
checked diagramatically.
In fact,  if $a = b$ holds the sum of ``external-leg correction" diagrams, 
picking up the effect of either $a$ or  $b$ at the first order of 
$m_{0l}^2$, just disappears (see Fig.3 ). Such cancellation mechanism does 
not seem to have been addressed  in
previous literature. Since right-handed charged lepton and
right-handed up-type quark super-multiplets both belong to 10-plet repr. of 
SU(5), if we  write $m_{0l}^2 = m_{0u}^2 = m_{10}^2$ and $m_{0h}^2 = 
m_{5}^2$,
and neglect the trilinear coupling, the condition quoted above reduces to 

\begin{equation}
10\cdot m_{10}^2 + 5\cdot m_{5}^2 = Str(
 M^2) = 0,
\end{equation}
where super-trace $Str$ is over a anomaly free set of matter multiplets.
This might suggest some theory which might have been hidden behind and
provides a scheme of natural lepton flavour conservation.

In SUGRA-inspired SUSY GUT, however, the above condition is not
chosen, and the logarithmically divergent LFV effect seems to remain.
Then situation is quite different from what we got for non-SUSY SU(5), where 
no logarithmic divergence remained and the decoupling theorem, expected
from a general argument, was valid. How should we interprete the remaining 
logarithmic divergence ?
The LFV SUSY breaking  slepton mass-squared   operator has $d = 2$ and 
should belong to the set of relevant operators, though we did not include it 
in $L_{rel}$ at tree level.
That is why the divergence appeared after the quantum correction. The 
divergence should be removed by the introduction of counterterms and is 
subject to a renormalization condition. Thus we are forced to make a 
conclusion that unfortunately
the logarithmic correction cannot be taken as a prediction of the theory, at 
least as far as we
work in the framework of SUSY GUT, as a renormalizable theory (calculable 
predictions of a theory should
all come from the finite quantum corrections to irrelevant operators).
Let us note that ignoring the LFV mass-squared  operators,
\begin{equation}
m_{LFV}^2 \ \tilde{e}_R^{\ast}\tilde{\mu}_R, \ \ \ m_{LFV}^2 
(|\tilde{\mu}_R|^2 - |\tilde{e}_R|^2),
\end{equation}
etc.,  does not enhance any symmetry of the theory, since SUSY has been 
explicitly broken by the flavour
independent SUSY breaking masses and also the flavour symmetry has been 
broken by the large top Yukawa coupling,
no matter the LFV mass-squared operators are present or not. In other words, 
there is no symmetry in the original lagrangian which guarantee small lepton 
flaovur violation.
We thus might have to say that \\
{\it ``the theory does not lead to natural suppression of lepton flavour 
violation."}  \\

One may wonder the situation discussed in Ref.\cite{Barbieri} that sleptom 
masses, whose boundary condition is set to be
flavour independent at $\Lambda = M_{pl}$, get logarithmic LFV corrections 
by the renormalization group effect is similar to the well-known evolution 
of gauge couplings
in ordinary SU(5) GUT, where three gauge couplings for SU(3), SU(2) and 
U(1), set to be all equal at $M_{GUT}$,
 deviate from each other in lower energies by renormalization group effect. 
In the case of the evolution of gauge couplings, however, the universal 
coupling at $M_{GUT}$
is naturally guaranteed by the symmetry of the theory, i.e. by SU(5).
Thus the splitting of gauge couplings comes not from the correction to the 
$L_{rel}$ but from the appearance of a new irrelevant operator with adjoint 
Higgs
field included. Thus the splitting is
genuine predicition of the theory, and is independent of the choice of the 
cutoff $\Lambda$.
On the other hand, in the case under consideration there will be no reason 
to expect that all SUSY breaking slepton masses evolve equally above 
$M_{pl}$ (, though we are not sure what the "above $M_{pl}$" means), even if 
they are
once unified at $M_{pl}$. We should note that flavour symmetry is hardly 
broken
 by large top Yukawa coupling, while SU(5) symmetry is softly broken by
the VEV of the adjoint Higgs in the issue of the gauge coupling evolution.

Finally we will ask further a question whether there is some chance to get 
the
logarithmic non-decoupling  LFV effect as a natural prediction of some 
theory.
More specifically, we will consider a renormalizable theory where the 
standard model
is included into an ``observable sector" of supersymetric GUT, and 
spontaneous SUSY breaking is realized in
a ``hidden" sector. These two sectors are assumed to be almost deoupled, 
being indirectly connected only by some interaction which acts on the both 
sectors and is assumed not to transmit the SUSY breaking to the SUSY GUT 
sector at the tree
level. We might think of
a ``gauge mediated SUSY breaking model",  for instance. At first glance the 
situation is
very similar to that of
the SUGRA theory with hidden sector, except that in SUGRA the SUSY breaking 
is transmitted to the observable sector already at the tree level via
non-renormalizable (super-)gravitational interaction. However we will not 
get
the non-decoupling effect in the renormalizable theory. The reason is that 
we cannot expect to get a log-divergent quantum correction to the LFV 
slepton masses in this case.
The absence of the LFV slepton masses in the lagrangian enhances 
supersymmetry, in contrast to the case of SUGRA where the absence does not 
enhance SUSY as there are universal SUSY breaking masses already at the tree 
lagrangian.  Thus as long as the theory is renormalizable, there will not 
appear any log-divergent correction to the masses. (The spontaneous SUSY 
breaking at the hidden sector will not affect the divergence.) The LFV 
slepton masses should be described by a higher dimensional ($d 
>4$)irrelevant operator whose coefficient is finite. The operator should 
necessarily
include $F$, an auxiliary component of some chiral field, as it is
caused by the spontaneous SUSY breaking due to $<F>$. As $d >4$, the 
coefficint should behave as
$M_{GUT}^{d-4}$, provided $M_{GUT}$ is the largest scale in the loop 
integral, since to get the LFV masses the GUT interaction is needed. Thus 
the LFV masses are expected to be suppressed at least by 
$\frac{M_{SUSY}^2}
{M_{GUT}^2}$, with $M_{SUSY}$ being SUSY breaking mass scale coming from 
$<F>$. Hence the rate of resultant $\mu \rightarrow e\gamma$ is anticipated 
to be
outrageously suppressed. It is interesting to note that the condition, 
Eq.(15), to cancel the log-divergence is trivially satisfied in this type of 
theories, as SUSY breaking masses, $m_{0l}^2$ etc., are all absent at the 
tree level.

\vspace{0.5 cm}
\leftline{\bf 4. Summary and discussions}
\vspace{0.2 cm}

After introductory and general (to some extent) argument on the mechanism of 
U.V. divergence cancellation in flavour changing neutral current processes 
and on the decoupling of particles with GUT scale masses, lepton-flavour 
violating processes, such as $\mu \rightarrow e\gamma$, were studied in 
ordinary (non-SUSY) SU(5) and SUSY SU(5). We have seen that in the 
 amplitude of $\mu \rightarrow e\gamma$ calculated in ordinary SU(5) 
logarithmic divergence cancels among  diagrams and remaining finite part are 
all suppressed by  at least $1/M_{GUT}^2$, in accordance with
the general argument. In SUSY SU(5), flavour changing slepton mass-squared 
 term does get a logarithmic correction, as claimed in the pioneering work 
\cite{Barbieri}. However, we have stressed the importance of the effect of 
flavour changing wave function renormalization. It turned out that when such 
effect is also taken into account  the
logarithmic correction disappears, provided a condition is met among SUSY 
 breaking soft masses. In SUGRA-inspired SUSY GUT, the condition is not 
satisfied.
The remaining logarithmic effect, however,  is argued not to be taken as a 
prediction of the theory associated with an irrelevant operator, in contrast 
to the case of the well-known splitting of three gauge couplings below 
$M_{GUT}$ by renormalization group effect in SU(5). We find that
 the log-correction should not exist in the renormalizable models with gauge 
mediated SUSY
breaking. It will be very challenging to search for a GUT-type  model which 
has calculable sizable amplitudes of lepton flavour violating processes as a 
natural prediction  of the theory, escaping the decoupling theorem.

\noindent {\bf Acknowledgment}

 The author would like to thank G. Branco, L. Lavoura and M.N. Rebelo for 
the very nice organization of the autumn school and warm hospitality during 
his stay in Lisboa. He is very grateful to T. Muta and T.  Morozumi for 
their  kind support by use of the international collaboration project of 
Ministry of Education, Japan, No. 08044089, which made this visit 
possible.

 The author also would like to thank F.J. Botella, T. Inami and B.Taga for 
very fruitful collaborations and valuable discussions in each stage, which 
this paper
is extensively based upon. Thanks are also due to N. Haba for  useful and 
informative arguments on the subject of gauge mediated SUSY breaking 
models.

 This work has been supported by Grant-in-Aid for Scientific Research 
(09640361) from the Ministry of Education, Science and Calture, Japan.

\end{document}